\documentclass[
    ,final            
  ]
  {aipproc}

\layoutstyle{6x9}

\usepackage{relsize}

\def\babar{\mbox{\slshape B\kern-0.1em{\smaller A}\kern-0.1em
    B\kern-0.1em{\smaller A\kern-0.2em R}}}
\def\belle{Belle}

\def\rhobar{\ensuremath{\overline{\rho}}}
\def\etabar{\ensuremath{\overline{\eta}}}
\def\sintwob{\ensuremath{\sin{2\beta}}}
\def\costwob{\ensuremath{\cos{2\beta}}}
\def\sintwobeff{\ensuremath{\sin{2\beta_\mathrm{eff}}}}
\def\CP{{\ensuremath{CP}}}
\def\ra{\ensuremath{\rightarrow}}

\usepackage{multirow}

\usepackage{feynmf}

\usepackage{maybemath}

\usepackage[italic]{hepparticles}

\usepackage{heppennames}

\begin{document}

SLAC-PUB-11557
\vspace{-1em}

\title{Measurements of the CKM Angle \boldmath{$\beta$}}



%
%

\author{Rainer Bartoldus}{
  address={
  Stanford Linear Accelerator Center\\
  2575 Sand Hill Rd\\
  Menlo Park, CA 90025},
  email={bartoldu@slac.stanford.edu},
}

%
%

\begin{abstract}
In this article I report on new and updated measurements of the
\CP-violating parameter $\beta$~($\phi_1$), which is related to the
phase of the Cabibbo-Kobayashi-Maskawa (CKM) quark-mixing matrix of
the electroweak interaction.
Over the past few years, $\beta$ has become the most precisely known
parameter of the CKM unitarity triangle that governs the \PB\ system.
The results presented here were produced by the two \PB\ Factories,
\babar\ and \belle, based on their most recent datasets of over 600
million $\PB\PaB$ events combined.
The new world average for \sintwob, measured in the theoretically and
experimentally cleanest charmonium modes, such as $\PBz\ra\PJgy\PKzS$,
is $\sintwob = 0.685 \pm 0.032$.
In addition to these tree-level dominated decays, independent
measurements of \sintwob\ are obtained from gluonic $\Pqb\ra\Pqs$
penguin decays, including $\PBz\ra\Pgf\PKzS$, $\PBz\ra\Pgh'\PKzS$ and
others.
There are hints, albeit somewhat weaker than earlier this year, that
these measurements tend to come out low compared to the charmonium
average, giving rise to the tantalizing possibility that New Physics
amplitudes could be contributing to the corresponding loop diagrams.
Clearly, more data from both experiments are needed to elucidate these
intriguing differences.
\end{abstract}

\maketitle


\section{The CKM Matrix}

Flavor transitions in the quark sector of the Standard Model are
described by the Cabibbo-Kobayashi-Maskawa (CKM) matrix, which links
the weak eigenstates of the three generations to their mass
eigenstates.
To maintain universality of the total coupling strength, the CKM
matrix must be unitary.
This leaves it with only four free parameters: three rotation angles
and one complex phase.
All other phases can be absorbed in the quark fields.
In the Standard Model, \CP\ violation is generated by this one
irreducible phase of the CKM matrix.

\subsubsection{The \PB\ Unitarity Triangle}

The unitarity condition leads to six relations of the form
$
V_{ud}V_{ub}^* + 
V_{cd}V_{cb}^* + 
V_{td}V_{tb}^* = 0
$,
which are geometrically represented as triangles in the complex plane.
They all have equal area, but only two of them have sides of the same
order, and thus naturally large angles.
Of these, the above unitarity relation is the one that controls \PB\ decays
and $\PB_d\PaB_d$ mixing.
The corresponding triangle is conveniently represented in the
parameters $\rhobar$ and
$\etabar$~\cite{Wolfenstein:1983yz}\cite{Buras:1994ec}, which give the
location of its apex
(see figure~\ref{fig:rhoeta_withs2abgam}).
The angle at the far point on the real axis is
$ \beta = \arg(-\frac{V_{cd}V_{cb}^*}{V_{td}V_{tb}^*})$.
%
A non-zero value of $\beta$ implies \CP\ violation.

\subsubsection{$\PB\PaB$ Mixing}

The second order weak interaction that causes the \PB\ and \PaB\ flavor
eigenstates to oscillate between each other is described by a pair of
box diagrams involving $W$ and top exchange.
(Other quarks are CKM-suppressed.)
As in the neutral kaon system, the \PB\ mass eigenstates are
superpositions of the flavor eigenstates:
$\PB_\mathrm{H} = p\, |\PB\rangle + q\, |\PaB\rangle$ and
$\PB_\mathrm{L} = p\, |\PB\rangle - q\, |\PaB\rangle$.
The oscillation frequency is given by the mass difference of the
heavy and the light states,
$\Delta m_\Pqd = m(\PB_\mathrm{H}) - m(\PB_\mathrm{L})$.
The box diagrams involve the CKM element $V_{td}$, which can to good
approximation be written as
$
V_{td} \approx |V_{td}|\, \mathrm{e}^{-i\beta}.
$
Thus, via mixing, the \PB\ meson picks up a weak phase of
$ \frac{q}{p} = \mathrm{e}^{-2i\beta}. $

\section{Time-Dependent \CP\ Asymmetries}

\subsubsection{Coherent Production of $\PB\PaB$}

It is worth recalling that in $\PgUc\ra\PB\PaB$ there is -- at any time --
exactly one $\PB_\mathrm{H}$ and one $\PB_\mathrm{L}$ (mass
eigenstates), one \PB\ and one \PaB\ (flavor eigenstates), as well as
one $\PB_{\CP=+1}$ and one $\PB_{\CP=-1}$ (\CP\ eigenstates).
It is this coherency that enables one to use the flavor tagging
techniques to see the interference between the two \PB\ mesons in the
event.

\subsubsection{Interference between Mixing and Decay}

As \CP\ eigenstates can be reached by the decay of both the \PB\ and
the \PaB, there is interference between decays with and without
mixing.
For a final state $f_\pm$, where $\pm$ denotes the flavor of the
decaying \PB\ to be \PB\ or \PaB, the decay rate as a function of the
\PB\ proper time, $\Delta t$, can be written as
\[
f_\pm(\Delta t) = \frac{1}{4 \tau_\PB}\mathrm{e}^\frac{-|\Delta t|}{\tau_\PB}
[ 1 \pm S_f \sin(\Delta m_d\Delta t) \mp C_f \cos(\Delta m_d\Delta t) ].
\]
Here, $\tau_\PB$ is the \PBz\ lifetime, and the coefficients $S_f$ and
$C_f$ for the sine and cosine term are
\[
S_f = \frac{2\,\mathrm{Im}\,\lambda}{1 + |\lambda|^2}, \qquad
C_f = \frac{1 - |\lambda|^2}{1 + |\lambda|^2},
\]
where
\[
\lambda = \frac{q}{p}\frac{\overline{A}(\PaB\ra f_\CP)}{A(\PB\ra f_\CP)}
\]
depends on the amplitude ratio of the \PaB\ and the \PB\ decay.
It is the (surprisingly) long \PB\ lifetime, which is comparable to
the oscillation frequency,
$(1/\tau_B \approx 0.5\,\mathrm{ps}^{-1}, \Delta m_\Pqd \approx
1.5\,\mathrm{ps}^{-1})$
that makes mixing observable.
The sine term ($S_f$) arises from the interference between direct
decay and decay after one net \PB-\PaB\ oscillation.
A non-zero cosine term ($C_f$) would arise from the interference
between decay amplitudes with different weak and strong phases (direct
\CP\ violation) or from \CP\ violation in $\PB\PaB$ mixing.

\subsubsection{\CP\ Asymmetry}

Using this, one can compute the time-dependent \CP\ asymmetry, which
becomes
\[
A_\CP(t) \equiv \frac{
\Gamma(\PaB(t)\ra f_\CP) -
\Gamma( \PB(t)\ra f_\CP)
}{
\Gamma(\PaB(t)\ra f_\CP) +
\Gamma( \PB(t)\ra f_\CP)
}
= S_{f} \sin(\Delta m_d\Delta t) - C_{f} \cos(\Delta m_d\Delta t).
\]
Again, $S_f$ is non-zero if there is \CP\ violation in the
interference between decays with and without mixing.  A non-zero value
for $C_f$ implies direct \CP\ violation.
In the absence of additional amplitudes with different weak phases,
and observing that in the Standard Model \CP\ violation in mixing is
negligible, one has
 $ |\lambda| \approx 1 $
and, to an excellent approximation,
$ S_f = \mathrm{Im}\,\lambda $
and
$ C_f = 0 $.
Hence, in this situation, which is the case for the charmonium final
states, and to some approximation for the $\Pqb\ra\Pqs$ penguins, the
sine coefficient becomes
$ S_f = \sintwob $.

\section{The Asymmetric \PB\ Factories \babar\ and \belle}

The two \PB\ factory experiments, \babar\ at SLAC in the US, and
\belle\ at KEK in Japan, started operations practically at the same
time in 1999.
Both facilities have an asymmetric beam energy configuration, with
9.0~GeV (\Pem) on 3.1~GeV (\Pep) at PEP-II, and 8.5~GeV (\Pem) on
3.5~GeV (\Pep) at KEKB, which leads to an effective boost of the
\PgUc\ system along the beam axis of $\beta\gamma = 0.56$ and
$\beta\gamma = 0.43$, respectively.
This opens the possibility to reconstruct decay time differences
between the two \PB\ mesons in the event by measuring the
displacements of their decay vertices along the beam line.
The cross section for $\Pqb\Paqb$ production at the \PgUc\ resonance
is about 1~nb, leading to 1 million $\PB\PaB$ pairs per fb$^{-1}$.
To date, both facilities have (far) exceeded their design
luminosities, taking that amount of data in only 1-2 days.
This has resulted in unprecedented datasets for \PB\ physics and much
beyond.

\subsubsection{\PB\ Reconstruction}

The general strategy for the reconstruction of \PB\ events is to
exploit the kinematics of the $\Pep\Pem\ra\PgUc\ra\PB\PaB$ process, in
which \PB\ mesons are produced nearly at rest in the \PgUc\ %
center-of-mass (cm) frame.
Two virtually uncorrelated kinematic variables are used to select \PB\ %
candidates: the so-called beam-energy constrained mass,
$m_\mathrm{EC} \equiv \sqrt{(E_\mathrm{beam}^\mathrm{cm})^2
-(p_\PB^\mathrm{cm})^2}$,
and the energy difference,
$\Delta E \equiv E_\PB^\mathrm{cm} - E_\mathrm{beam}^\mathrm{cm}$,
where
$E_\mathrm{beam}^\mathrm{cm}$ is the beam energy in the \PgUc\ cm
frame, and $E_\PB^\mathrm{cm}$ and $p_\PB^\mathrm{cm}$ are the cm
energy and momentum of the \PB\ candidate, respectively.
%
%
Various multivariate techniques were developed that distinguish
$\PB\PaB$ events from ($\Pep\Pem\ra\Pq\Paq$, $\Pq =
\Pqu,\Pqd,\Pqs,\Pqc$) continuum as well as from potential QED
backgrounds such as Bhabha events.
Many methods take advantage of the fact that $\PB\PaB$ event shapes
tend to be spherical, whereas continuum background is more jet-like.

\subsubsection{Time-Dependent \CP\ Analysis }

To extract the $\Delta t$ distribution with high efficiency, both
experiments have developed sophisticated flavor tagging techniques.
One \PB\ meson is fully reconstructed in a \CP\ eigenstate, which also
determines its decay vertex.
The other \PB\ is not reconstructed, but its flavor is determined
("tagged") to be either a \PB\ or a \PaB, from one of various tagging
algorithms.
These include, for example, lepton tags from semi-leptonic decays,
kaon tags, soft pion tags from \PDstpm\ decays, etc.
The tags are defined using multivariate algorithms, involving
likelihood selectors or neural networks.
A mistag probability, $w$, dilutes the observed asymmetry -- and
reduces the sine amplitude -- by a factor $(1-2w)$.
Thus, a figure of merit for the \CP\ analysis is the effective tagging
efficiency,
$Q = \sum_i{\epsilon_i(1-2w_i)^2}$,
where $\epsilon_i$ is the tagging efficiency of mode $i$.
Both experiments achieve an effective tagging efficiency that is very
close to $30\,\%$.
The tagging vertex is determined from tracks not associated with the
reconstructed \PB\ candidate.
The measured separation between the two decay vertices, $\Delta z$,
gives $\Delta t = \Delta z / \beta\gamma$.
The time difference, $\Delta t$, is a signed quantity, as the
principle applies whether the tagged \PB\ decays before or after the
reconstructed \PB.
The average separation $\langle\Delta z\rangle$ in the two experiments
is about $100-200\,\mu\mathrm{m}$.

\subsubsection{Trees and Penguins}

\begin{figure}
\begin{fmffile}{fmfbeta}

\fmfframe(0,0)(100,20)
{
\begin{fmfgraph*}(120,60)
\fmfstraight
\fmfleftn{i}{2}
\fmflabel{$d$}{i1}
\fmflabel{$\bar b$}{i2}
\fmfrightn{o}{4}
\fmflabel{$d$}{o1}
\fmflabel{$\bar s$}{o2}
\fmflabel{$c$}{o3}
\fmflabel{$\bar c$}{o4}
\fmf{fermion}{o4,v1,i2}
\fmffreeze
\fmf{photon,label=$W$,l.side=right}{v1,v2}
\fmf{fermion}{o2,v2,o3}
\fmf{fermion}{i1,o1}
\fmf{phantom}{i1,v3,o1}
\fmf{phantom}{v3,v2}
\fmfdot{v1,v2}
\fmflabel{$V_{cb}^*$}{v1}
\fmfv{l=$V_{cs}$,l.a=90}{v2}
\fmf{phantom,label=\PBz,l.side=left,l.dist=.15w}{i1,i2}
\fmf{phantom,label=\PJgy,l.side=right,l.dist=.2w}{o3,o4}
\fmf{phantom,label=\PKz,l.side=right,l.dist=.2w}{o1,o2}
\end{fmfgraph*}
}

\fmfframe(0,100)(0,20)
{
\begin{fmfgraph*}(120,60)
\fmfstraight
\fmfleftn{i}{2}
\fmflabel{$d$}{i1}
\fmflabel{$\bar b$}{i2}
\fmfrightn{o}{4}
\fmflabel{$d$}{o1}
\fmflabel{$\bar s$}{o2}
\fmflabel{$s$}{o3}
\fmflabel{$\bar s$}{o4}
\fmf{fermion}{o4,v3}
\fmf{plain,tension=2}{v3,v2,v1}
\fmf{phantom,tension=0,label=$\bar t$,l.side=right}{v3,v1}
\fmf{fermion}{v1,i2}
\fmffreeze
\fmf{boson,left,label=$W$,l.side=left}{v1,v3}
\fmf{gluon,label=$g$,l.side=right,l.dist=.1w}{v2,v4}
\fmf{fermion}{o2,v4,o3}
\fmf{fermion}{i1,o1}
\fmf{phantom}{i1,v5,o1}
\fmf{phantom}{v5,v4}
\fmfdot{v1,v3}
\fmflabel{$V_{tb}^*$}{v1}
\fmflabel{$V_{ts}$}{v3}
\fmf{phantom,label=\PBz,l.side=left,l.dist=.15w}{i1,i2}
\fmf{phantom,label=\Pgf,l.side=right,l.dist=.2w}{o3,o4}
\fmf{phantom,label=\PKz,l.side=right,l.dist=.2w}{o1,o2}
\end{fmfgraph*}
}

\end{fmffile}

\label{fig:diagrams} \caption{Feynman diagrams for \PB\ decays proceeding
via tree (left) and penguin (right) transitions.}
\end{figure}
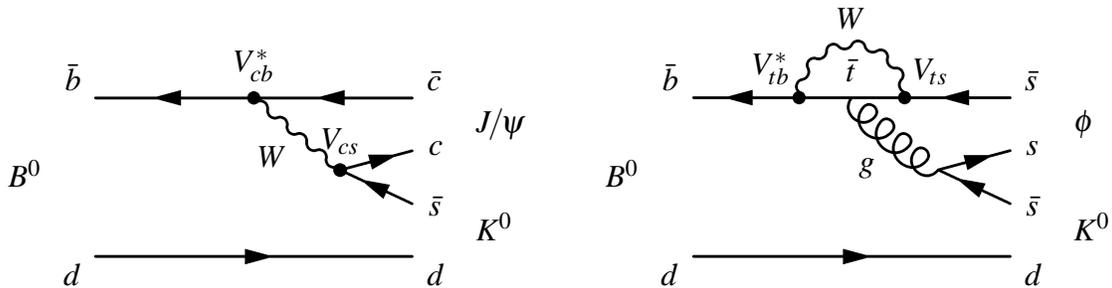

The $\Pqb\ra\Pqc\Paqc\Pqs$ transitions are mediated by tree and
penguin diagrams, with equal dominating weak phases.
The $\Pqb\ra\Pqs\Paqs\Pqs$ transitions are pure penguin diagrams.  Due
to the high mass scales involved in the penguin loops, new particles
could enter and contribute additional weak phases, thus giving rise to
\CP\ violation beyond the Standard Model.
See figure~\ref{fig:diagrams} for the two main diagrams.

\section{\boldmath{\sintwob} from \boldmath{$\PB\ra\Pqc\Paqc\PKz$} (charmonium)}

\begin{figure}
  \includegraphics[height=.5\textheight]{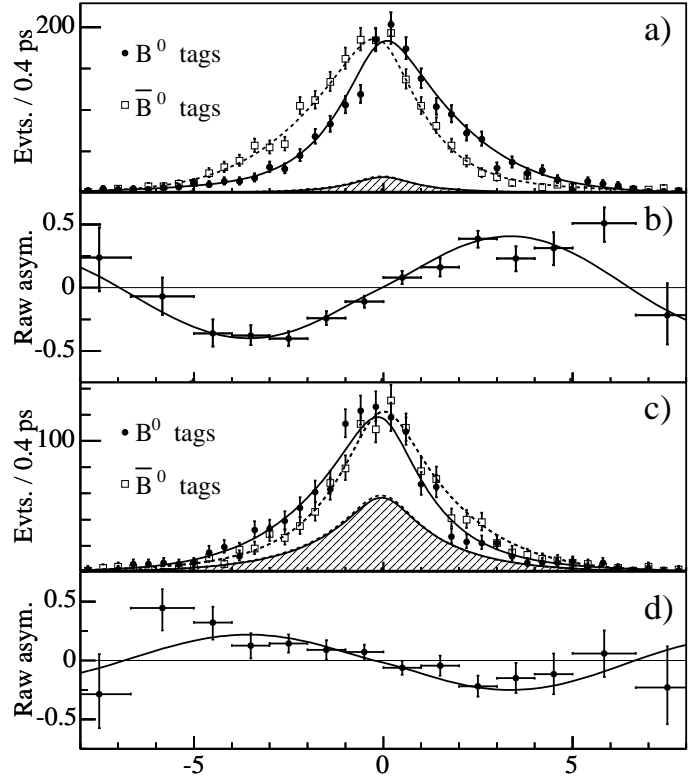}

  \caption{\babar\ $\Delta t$ distributions of candidates with \PBz\
  and \PaBz\ tags for \CP-odd (a) and \CP-even (c) modes, and
  corresponding raw asymmetries (b and c). The solid curves represent
  the projection from the maximum likelihood fit.}

 \label{fig:sin2beta_babar}
\end{figure}

The \CP\ asymmetries in the proper-time distribution of neutral \PB
decays into a charmonium and a \PKz\ meson provide a high-precision
measurement of \sintwob.
In their analysis of $227\times10^6\,\PB\PaB$ decays \babar\ exploits
all currently accessible final states: $\PJgy\PKzS$, $\PJgy\PKzL$,
$\Pgyii\PKzS$, $\Pcgc\PKzS$, $\Pcgh\PKzS$ and
$\PJgy\PKst^0(\PKst^0\ra\PKzS\Pgpz)$ \cite{Aubert:2004zt}.
Of these, the \PKzS\ modes are \CP-odd, the one including
\PKzL\ is \CP-even, and the $\PJgy\PKst^0$ state involves
contributions from either \CP\ state, where the effective eigenvalue
is computed from the relative fractions of odd and even orbital
angular momenta.
\babar\ performs a global maximum likelihood fit with 65 free
parameters, including -- besides \sintwob -- mistag fractions for all
tagging categories, $\Delta t$ resolution functions and time
dependence for signal and background samples.
From this, \babar\ obtains
\[
\sintwob = 0.722 \pm 0.040 \pm 0.023, \qquad\mbox{(\babar)}
\]
where, as throughout this note, the first error is statistical and the
second systematic.
Figure~\ref{fig:sin2beta_babar} shows the $\Delta t$ distributions and
raw asymmetries.

\begin{figure}
  \includegraphics[height=.4\textheight]{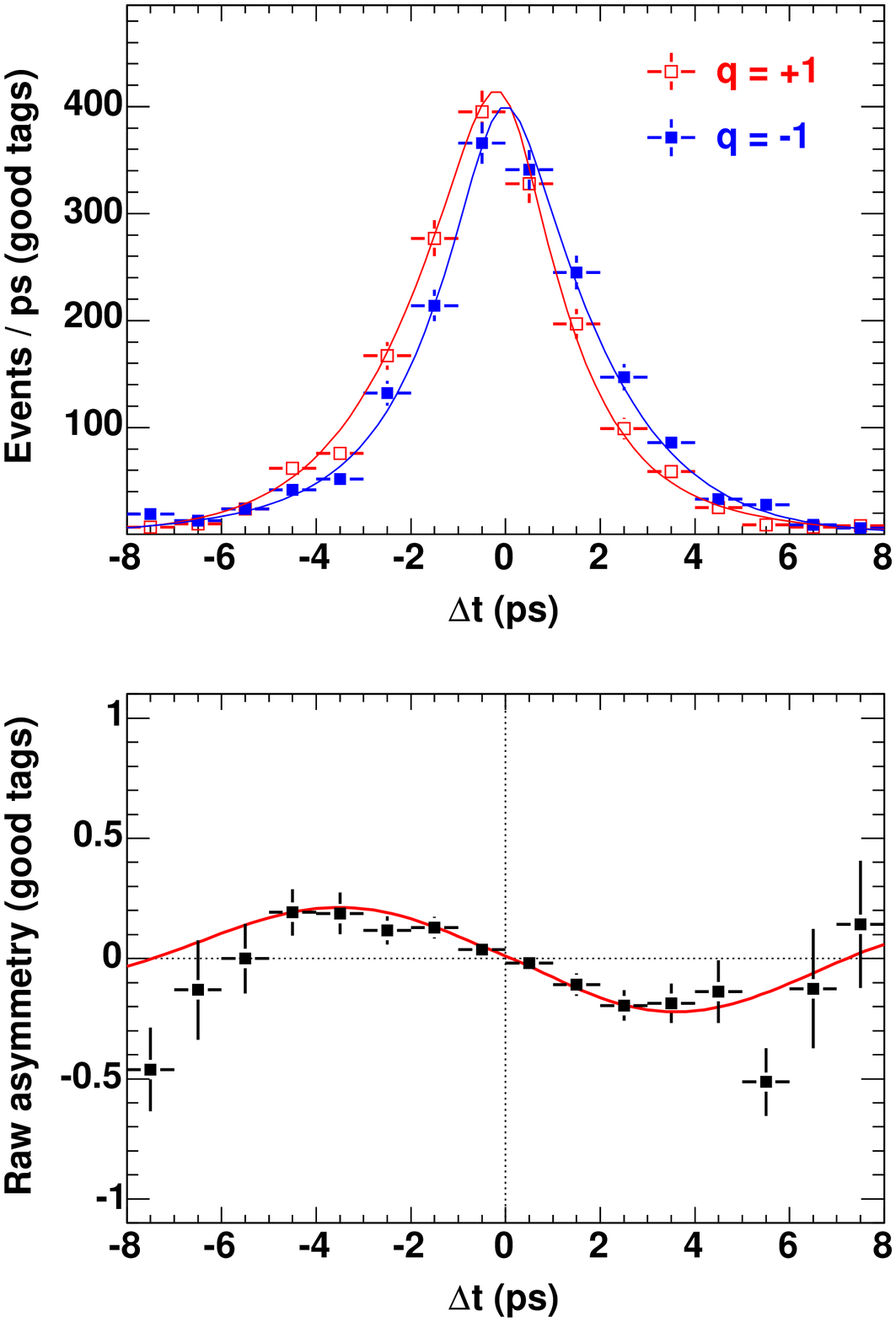}
  \includegraphics[height=.4\textheight]{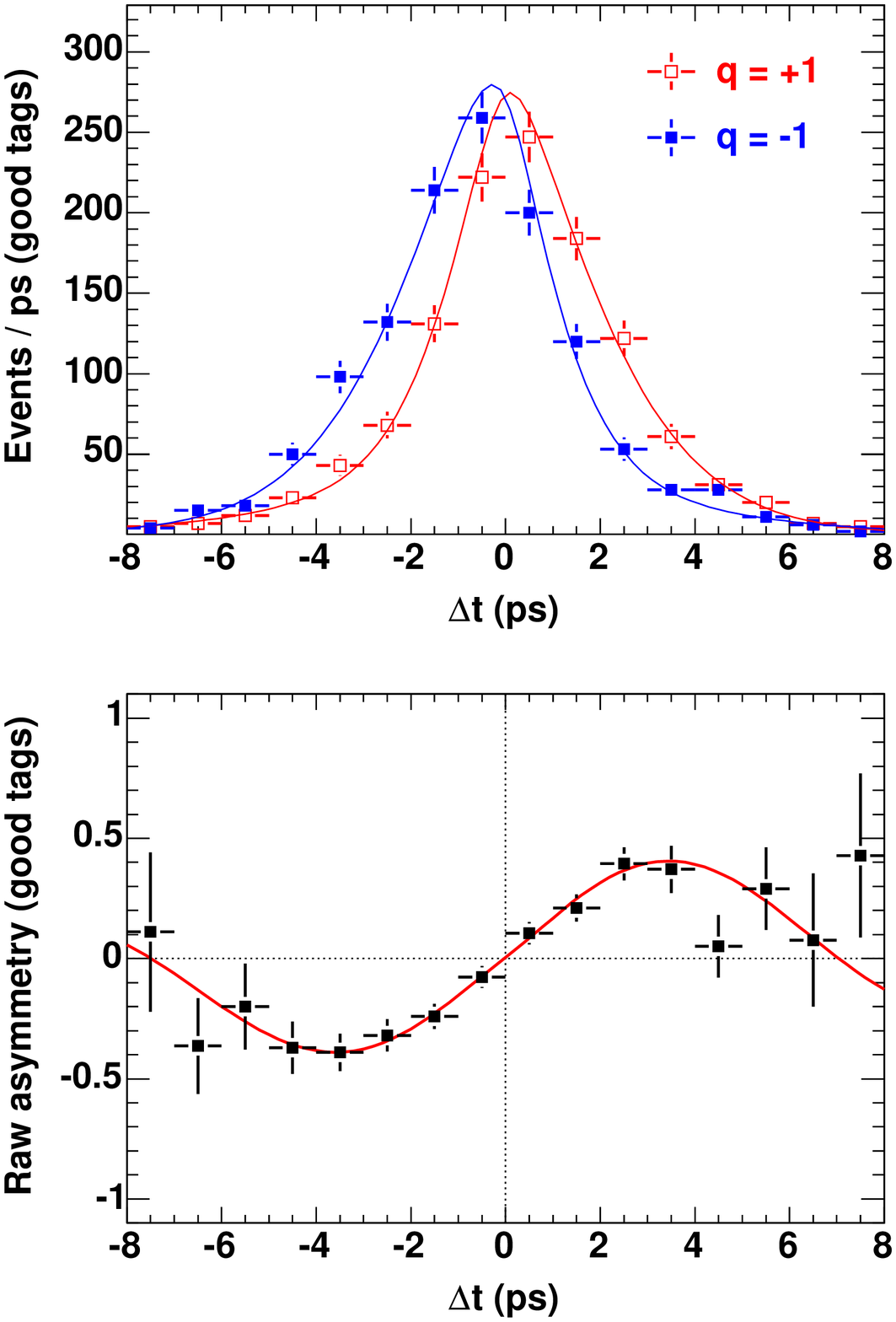}

  \caption{\belle\ $\Delta t$ distributions (top) in $\PJgy\PKzS$
  (left) and $\PJgy\PKzL$ (right) final states for \PBz\ and \PaBz
  tags, and corresponding raw asymmetries (bottom).  The data shown
  corresponds to "good" ($0.5 < r \le 1$) tags. The curves show the
  result of the maximum likelihood fit.}

  \label{fig:sin2beta_belle}
\end{figure}

The very recent \belle\ analysis is based on $386\times10^6\,\PB\PaB$
pairs, while focusing on the final states $\PJgy\PKzS$ and
$\PJgy\PKzL$ \cite{Abe:2005bt}.
\belle\ divides the data using an event-by-event flavor tagging
dilution factor $r$, determined from Monte Carlo, that varies from 0
for no flavor discrimination to 1 for unambiguous flavor assignment.
For the fit procedure, \belle\ takes a somewhat complementary approach
to \babar, where resolution functions and mistag fractions are
determined in a first, separate multiparameter fit to various control
samples, leaving only the two coefficients $S_f$ and $C_f$ as free
parameters in the final maximum likelihood fit.
The new \belle\ result is
\[
\sintwob = 0.652 \pm 0.039 \pm 0.020. \qquad\mbox{(\belle)}
\]
The $\Delta t$ distributions and raw asymmetries are shown in
figure~\ref{fig:sin2beta_belle}.

\subsubsection{UT Constraints in the $\rhobar$-$\etabar$ Plane}

\begin{figure}
  \includegraphics[width=.75\textwidth]{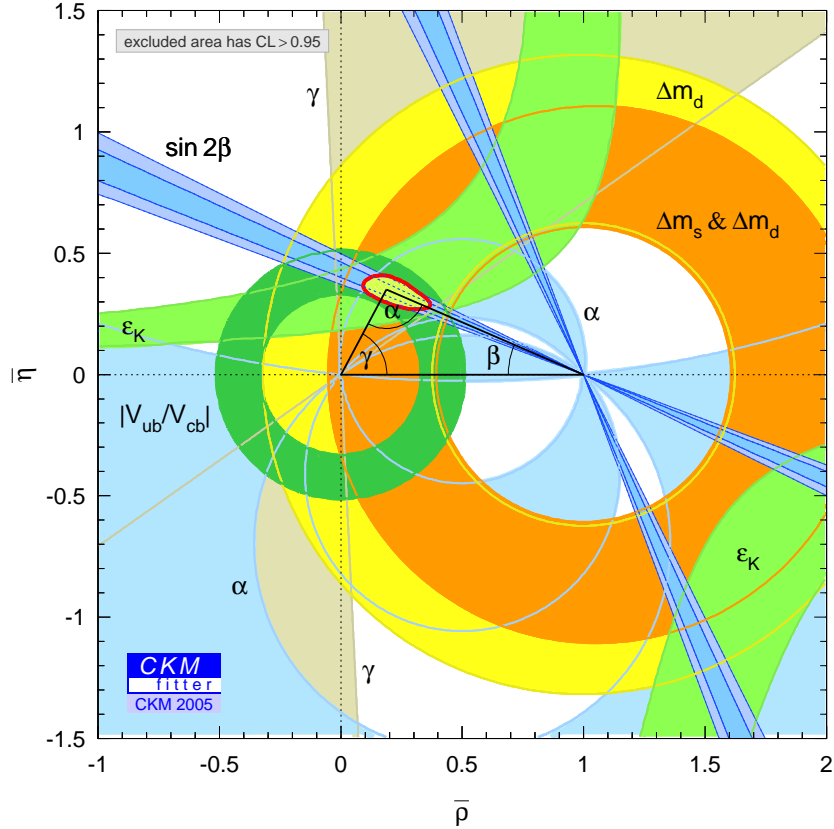}
  \caption{Constraints in the (\rhobar,\etabar) plane using all
  available inputs, including the most recent $\beta$ measurements, in
  the global CKM fit.}
  \label{fig:rhoeta_withs2abgam}
\end{figure}

Combining the results from the two experiments with all available
inputs is among the charges of the Heavy Flavor Averaging
Group~\cite{Group(HFAG):Triangle/Summer2005}.
The result for \sintwob\ is to be compared with information from other
measurements that provide constraints on the Unitarity Triangle.
This information can be neatly presented as allowed regions of various
shapes in the $\rhobar$-$\etabar$ plane.
Figure~\ref{fig:rhoeta_withs2abgam} shows the allowed regions from
\sintwob, from the other two UT angles, $\alpha$ and $\gamma$, from
the measurement of $|V_{\Pqu\Pqb}/V_{\Pqc\Pqb}|$, as well as from the
\PdB\ and \PsB\ mixing results for $\Delta m_\Pqd$ and $\Delta
m_\Pqs$, and from $\epsilon_\PK$ from \CP\ violation in the \PK
system.
It can be seen that the \sintwob\ results are in remarkable agreement
with all other UT measurements.
This shows that the CKM mechanism is indeed the main source of \CP\
violation.
Combining these constraints in a global fit, as performed by the
CKMfitter group~\cite{Charles:2004jd,Group(CKMfitter):LP2005}, leaves
only a small allowed region for the position of the
$(\rhobar,\etabar)$ apex.

\subsubsection{UT Constraints from Angles alone}

\begin{figure}
  \includegraphics[width=.75\textwidth]{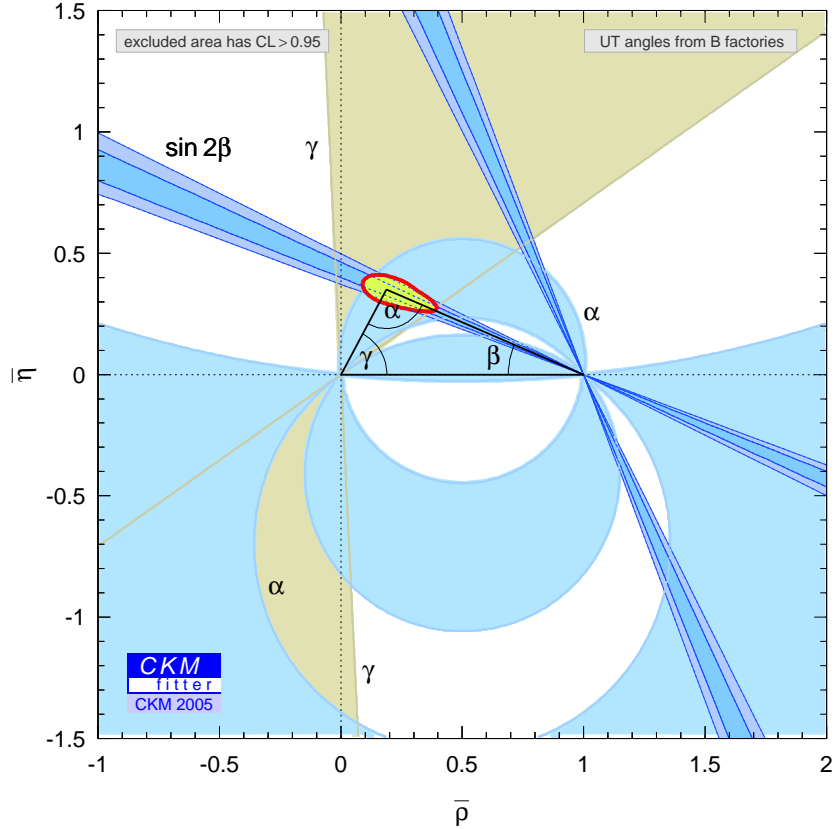}
  \caption{Constraints in the (\rhobar,\etabar) plane including only
  the angle measurements in the CKM fit.}
  \label{fig:rhoeta_angles}
\end{figure}

If one ignores the inputs from the UT sides, from \PB\ mixing and from
the \PK\ system, one is left with only the angle measurements,
\textit{i.e.}, measurements of \CP\ violation in the \PB\ system.
A combined fit to only these inputs returns an allowed
$\rhobar-\etabar$ region very close to that of the full fit,
demonstrating that today most of the constraint comes from the angles
alone (figure~\ref{fig:rhoeta_angles}).
This can be seen as a milestone for the \PB\ Factories and illustrates
how powerful they are.
Moreover, \sintwob\ is the first UT input that is not limited by
theory uncertainties.

It should also be noted that one can resolve the ambiguity in the sign
of \costwob\ by a time-dependent angular analysis of the vector-vector
final state $\PJgy\PKstz$, which disfavors the solution with the
larger \sintwob\ value at the $95\,\%$ confidence
level~\cite{Aubert:2004cp}\cite{Itoh:2005ks}.


\section{\boldmath{\sintwob} from \boldmath{$\Pqb\ra\Pq\Paq\Pqs$} Penguins}

The effective \sintwob\ values, \sintwobeff, from the sine coefficient
in penguin modes, are expected to agree with the charmonium values to
within a few percent.
Uncertainties are smallest for the pure penguin modes,
$\PBz\ra\Pgf\PKz$ and $\PBz\ra\PKzS\PKzS\PKzS$, and larger for the
other modes, due to a possible $\Pqb\ra\Pqu$ transition that carries a
weak phase $\gamma$.
Larger deviations would indicate a new \CP-violating weak phase beyond
the Standard Model.
As already mentioned, large virtual mass scales are involved in the
penguin loops, which may lead to additional diagrams with new heavy
particles.
A program is underway to measure as many of these modes as possible~
\cite{Aubert:2005ja,Aubert:2005vw,Aubert:2004gk,Aubert:2005zt,Aubert:2005ra,Aubert:2005ha,Aubert:2005kd}\cite{Abe:2005bt}.


\subsubsection{$\PBz\ra\Pgf\PKz$}

The decay mode $\PBz\ra\Pgf\PKz$ is dominated by the gluonic penguin
transition $\Pqb\ra\Pqs\Paqs\Pqs$.
The neutral kaon is reconstructed in the \CP-odd mode as \PKzS, from
$\PKzS\ra\Pgpp\Pgpm$ and $\PKzS\ra\Pgpz\Pgpz$, and in the \CP-even
mode as \PKzL, using calorimeter and muon chamber signatures
\cite{Aubert:2005ja,Abe:2005bt}.


\subsubsection{$\PBz\ra\PKp\PKm\PKzS$}

The decays $\PBz\ra\PKp\PKm\PKzS$, excluding the resonant $\Pgf\PKzS$
contribution, are in general not \CP\ eigenstates but rather an
admixture of \CP-even ($f_+$) and \CP-odd ($f_-$) components.
The \CP\ eigenvalue depends on the angular momentum of the $\PKp\PKm$
system: it is \CP-odd for a relative $P$-wave, and \CP-even for an
$S$-wave.
The observed sine coefficient therefore becomes $S_f = (2 f_+ - 1)\,
\sintwobeff$.
To obtain \sintwobeff, the fraction $f_+$ needs to be determined
experimentally.
\babar\ and \belle\ follow different approaches.
\babar\ performs an angular moment analysis based on the helicity
angle distribution of one of the charged \PK\ mesons to extract the
CP-even content as $f_+ = 0.89\pm0.08\pm0.06$
\cite{Aubert:2005ja}.
\belle\ also finds the state to be predominantly CP-even by using an
isospin relation which yields $f_+ = 0.93\pm0.09\pm0.05$
\cite{Abe:2005bt}.


\subsubsection{$\PBz\ra\Pgh'\PKz$}

The $\PBz\ra\Pgh'\PKz$ decay is another theoretically clean mode.
It also enters with the highest branching fraction of all
$\Pqb\ra\Pqs$ penguin modes being investigated.
The \Pghpr\ meson is reconstructed in the decays $\Pghpr\ra\Pgr^0\Pgg$
and $\Pghpr\ra\Pgh\Pgpp\Pgpm$ with $\Pgh\ra\Pgg\Pgg$ or
$\Pgh\ra\Pgpp\Pgpm\Pgpz$.
\babar\ reconstructs the \PKzS, \belle\ also includes the \PKzL\
\cite{Aubert:2005vw,Abe:2005bt}.


\subsubsection{$\PBz\ra\Pgpz\PKzS$}

The mode $\PBz\ra\Pgpz\PKzS$ obviously poses a challenge for the
reconstruction of the \CP\ decay vertex.
A method has been developed that exploits the knowledge of the
average interaction point (IP), which is determined on a run-by-run
basis from the spatial distribution of two-prong events.
By constraining the single \PKzS\ trajectory to this IP, $\Delta t$ is
computed in a geometric fit.
The sensitivity is further improved by constraining the sum of the two
\PB\ decay times to $2\,\tau_\PB$
\cite{Aubert:2005zt,Abe:2005bt}.


\subsubsection{$\PBz\ra\PKzS\PKzS\PKzS$}

With no charged track coming from the IP, the $\PBz\ra\PKzS\PKzS\PKzS$
decay vertex has to be reconstructed using the same IP-constraint
method pioneered for the mode $\Pgpz\PKzS$.
At least two \PKzS\ are reconstructed in $\PKzS\ra\Pgpp\Pgpm$,
allowing one to be $\PKzS\ra\Pgpz\Pgpz$.
Theoretically, the $\PBz\ra\PKzS\PKzS\PKzS$ mode is as clean as
$\PBz\ra\Pgf\PKz$, since there is no \Pqu\ quark in the final state,
and it is dominated by the $\Pqb\ra\Pqs\Paqs\Pqs$ penguin transition
\cite{Aubert:2005kd,Abe:2005bt}.

\subsubsection{Summary}

\begin{figure}
  \includegraphics[height=.5\textheight]{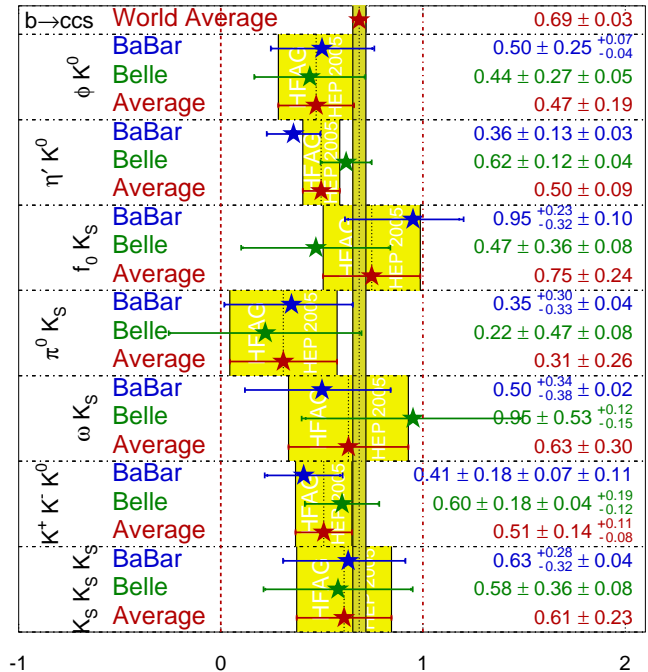}

  \caption{Compilation of results for \sintwob\ from charmonium and
  \Pqs penguin decays.}
  \label{fig:sPengS_CP}
  
\end{figure}

Table~\ref{tab:yields} summarizes the selected \sintwob\ modes with
their \CP\ values, branching fractions and the event yields obtained
by the two experiments.
\begin{table}
\begin{tabular}{lccrr}
\hline
    \tablehead{1}{l}{b}{Mode}
  & \tablehead{1}{c}{b}{\CP}
  & \tablehead{1}{c}{b}{branching \\ fraction $(10^{-5})$}
  & \tablehead{1}{r}{b}{\babar\ \\ signal yield}
  & \tablehead{1}{r}{b}{\belle\ \\ signal yield}
  \\
\hline
$\PJgy\PKzS$      & $-1$
  & \multirow{2}{*}{$85 \pm 5$}
  & $3404$
  & $5264 \pm 73$
  \\
$\PJgy\PKzL$      & $+1$
  &
  & $2788$
  & $4792 \pm 105$
  \\
$\Pgyii\PKzS$     & $-1$
  & $62 \pm 7$
  & $485$
  &
  \\
$\Pcgc\PKzS$      & $-1$
  & $40 \pm 12$
  & $194$
  &
  \\
$\Pcgh\PKzS$      & $-1$
  & $116 \pm 26$
  & $287$
  &
  \\
$\PJgy\PKst^0$     & $-1$
  & $131 \pm 7$
  & $572$
  & 
  \\
\hline
$\Pgf\PKzS$       & $-1$ 
  & \multirow{2}{*}{$0.86^{+0.13}_{-0.11}$}
  & $114 \pm 12$
  & $180 \pm 16$
  \\
$\Pgf\PKzL$       & $+1$ 
  & $$
  & $98 \pm 18$
  & $78 \pm 13$
  \\
$\Pgh'\PKzS$      & $-1$ 
  & \multirow{2}{*}{$6.3 \pm 0.7$}
  & $804 \pm 40$
  & $830 \pm 35$
  \\
$\Pgh'\PKzL$      & $+1$ 
  &
  &
  & $187 \pm 18$
  \\
$\PKzS\PKzS\PKzS$ & $+1$ 
 & $0.42^{+0.18}_{-0.15}$
  & $ 88 \pm 10$
  & $105 \pm 12$
  \\
$\Pgpz\PKzS$      & $-1$ 
  & $1.2 \pm 0.1$
  & $186 \pm 18$ 
  & $106 \pm 14$ 
  \\
$\Pfz\PKzS$       & $+1$ 
  & $$
  & $152 \pm 19$
  & $145 \pm 16$
  \\
$\Pgo\PKzS$       & $-1$ 
  & $0.55^{+0.12}_{-0.10}$
  & $92 \pm 13$
  & $68 \pm 13$
  \\
\multirow{2}{*}{$\PKp\PKm\PKzS$}   &  $0.89\pm0.08\pm0.06$
  & \multirow{2}{*}{$2.5 \pm 0.2$}
  & \multirow{2}{*}{$452 \pm 28$}
  & \multirow{2}{*}{$536 \pm 29$}
  \\
                                   &  $0.93\pm0.09\pm0.05$
  &
  &
  &
  \\
\hline
\end{tabular}

\caption{Signal yields for the time-dependent \CP\ analysis selection
of all tree and penguin \PB\ decay modes.}
\label{tab:yields}
\end{table}
Figure~\ref{fig:sPengS_CP} summarizes the \sintwob\ results from the
gluonic penguin modes, adding averages between \babar\ and \belle, in
comparison with the new charmonium world average.















\section{Conclusions}

In this article I briefly reviewed measurements of $\beta$~($\phi_1$)
that were obtained by the \PB\ factory experiments, \babar\ and
\belle, from the analysis of two different sources: the theoretically
and experimentally "golden" \PB\ decay modes into charmonium final
states, $\Pqb\ra\Pqc\Paqc\Pqs$, and a number of gluonic penguin modes
involving the transition $\Pqb\ra\Pqs\Paqs\Pqs$.
The charmonium modes, having provided first evidence of \CP\ violation
outside the neutral \PK\ system only a few years ago, have now reached
a precision of better than $5\,\%$ from both experiments combined.
The remarkable agreement of \sintwob\ with other Unitarity Triangle
constraints establishes the CKM mechanism as the dominant source of
\CP\ violation.
At the same time it makes it a firm reference for SM tests.
Corresponding measurements of time-dependent \CP\ violation in
$\Pqb\ra\Pqs$ penguins, on the other hand, seem to reveal consistently
lower values for \sintwob, thus leaving room for possible
contributions of New Physics that could enter the loops.
Increasingly sophisticated analyses of these rather challenging modes
on one side, and better theoretical calculations on the other, have
helped shape our knowledge surrounding this apparent difference.
At this point we cannot know if the discrepancy is going to go away
-- as all others have in the history of the Standard Model -- or
whether it might indeed reveal the first hint of New Physics.
We shall look forward to more results from the \PB\ factories, who in
their continued operation each expect to quadruple the statistics of
their datasets in the coming three years.


\begin{theacknowledgments}
I like to thank my \babar\ colleagues who provided me with input for
this talk.
Particular thanks go to Yoshi Sakai from \belle\ for giving me access
to the yet unfinished draft of the \sintwob\ conference preprint with
the newest measurements.
Receiving all new results just days before the conference definitely
made this an exciting trip.
Finally, I want to say special thanks to my friend Andreas H{\"o}cker,
who gave me insight into the yet preliminary HFAG fits and averages,
as well as the nice CKMfitter results and plots.
As always, I truly enjoyed our stimulating discussions about CP
violation, \PB\ physics, and everything else.
\end{theacknowledgments}


\bibliographystyle{aipproc}   

\bibliography{slac-pub-11557}

\end{document}